\documentclass[12pt,twoside,english,fleqn]{article}
\usepackage[T1]{fontenc}
\usepackage[latin9]{inputenc}
\usepackage{geometry}
\usepackage{color}
\usepackage{babel}
\usepackage{units}
\usepackage{amsmath}
\usepackage{amssymb}

\usepackage{multirow}
\usepackage[authoryear]{natbib}
\usepackage[unicode=true,
 bookmarks=true,bookmarksnumbered=true,bookmarksopen=true,bookmarksopenlevel=1,
 breaklinks=false,pdfborder={0 0 1},backref=false,colorlinks=true]
 {hyperref}
\usepackage{breakurl}

\usepackage{graphicx}

\makeatletter

\providecommand{\tabularnewline}{\\}

\newenvironment{lyxlist}[1]
{\begin{list}{}
{\settowidth{\labelwidth}{#1}
 \setlength{\leftmargin}{\labelwidth}
 \addtolength{\leftmargin}{\labelsep}
 }}
{\end{list}}

\usepackage{babel}
\usepackage{babel}
\@ifundefined{definecolor}{color}{}
\@ifundefined{definecolor}{color}{}
\@ifundefined{definecolor}{color}{}

\usepackage{longtable}
\usepackage{ifpdf}
\ifpdf 

 \IfFileExists{lmodern.sty}
  {\usepackage{lmodern}}{}

\fi 

\let\myTOC\tableofcontents
\renewcommand{\tableofcontents}{%
  \frontmatter
  \pdfbookmark[1]{\contentsname}{}
  \myTOC
  \mainmatter }

\def\LyX{\texorpdfstring{%
  L\kern-.1667em\lower.25em\hbox{Y}\kern-.125emX\@}
  {LyX}}

\usepackage{multicol}

\makeatother

\title{A new hierarchy of phylogenetic models consistent with heterogeneous substitution rates}

\author{Michael D. Woodhams, Jes\'us Fern\'andez-S\'anchez, Jeremy G. Sumner}

\begin{document}

\maketitle

\begin{abstract}
\noindent
When the process underlying DNA substitutions varies across evolutionary history, the standard Markov models underlying standard phylogenetic methods are mathematically inconsistent. 
The most prominent example is the general time reversible model (GTR) together with some, but not all, of its submodels.
To rectify this deficiency, Lie Markov models have been developed as the class of models that are consistent in the face of a changing process of DNA substitutions.
Some well-known models in popular use are within this class, but are either overly simplistic (e.g. the Kimura two-parameter model) or overly complex (the general Markov model). 
On a diverse set of biological data sets, we test a hierarchy  of Lie Markov models spanning the full range of parameter richness. 
Compared against the benchmark of the ever-popular GTR model, we find that as a whole the Lie Markov models perform remarkably well, with the best performing models having eight parameters and the ability to recognise the distinction between purines and pyrimidines. \\


\end{abstract}

\vfill
\hrule\mbox{}\\
\thanks{\footnotesize{
\noindent
MD Woodhams $\cdot$ JG Sumner \\ 
School of Mathematics and Physics, University of Tasmania, Tasmania, Australia\\

\noindent
J Fern\'andez-S\'anchez\\
Departament de Matem\`atica Aplicada I, Universitat Polit\`ecnica de Catalunya, Barcelona, Spain\\

\noindent
email: michael.woodhams@utas.edu.au, jesus.fernandez.sanchez@upc.edu, jsumner@utas.edu.au
}
}

\section{Introduction and Motivation}

\noindent 
Exclusively from a mathematical point of view, \citet{LieMarkov} introduced the Lie Markov models of DNA evolution which have the property of closure under matrix multiplication. 
In  section~\ref{sec:construction}, we will give a detailed explanation of what is meant by closure and why it is of practical importance, but essentially it ensures that an inhomogeneous process (where rate matrices from a particular model change with time) is equivalent to an ``average'' homogeneous process (using rate matrices obtainable from the same model).
Models which do not have this property (notably including GTR \citep{LieMarkov}) have
a consistency problem when modeling an inhomogeneous process: if a
sequence evolves for a time under one set of GTR rate parameters,
then for a time under a different set of GTR rate parameters, the
joint probabilities (pattern frequencies) between the start and end
of this process cannot (in general) be described by a single GTR model.
One consequence of this is that, in an inhomogeneous GTR model (i.e.
different GTR Markov matrices on each branch of a tree), pruning the
tree changes the distribution of site patterns achievable at the remaining taxa.
Thus a ``closed'' model can be defined in the narrow sense that
the Markov matrices are closed under matrix multiplication, but also
in a broader sense in which the corresponding phylogenetic model (as
a set of candidate site pattern distributions) is ``closed''
under pruning of the tree (via marginalization), with the former implying the latter. The practical
significance of model misspecification that can occur when implementing
a model that is not closed under matrix multiplication has been explored
by \citet{sumner2012b}.

\citet{LieMarkov} derived the hierarchy of Lie Markov models with
maximal symmetry (those that treat all nucleotides equivalently).
This hierarchy consists of the Jukes-Cantor (one-parameter) model
\citep{JC}, the K3ST (three-parameter) model \citep{K3P}, the F81
(four-parameter) model \citep{F81}, the general Markov (twelve-parameter)
model \citep{GMM}, and a previously unknown six-parameter model ``F81+K3ST'',
which has rate matrices that are the sum of F81 and K3ST rate matrices.
In \citet[Table 2]{sumner2012b}, these models were compared to GTR under an Akaike Information Criterion \citet{AIC} framework.
There it was found that F81+K3ST was marginally superior to GTR on one data set
(human mitochondrial genomes), and markedly inferior on the other
four data sets examined. 
Despite its novelty, a practical disadvantage of the F81+K3ST
model is that it does not account for the biological fact that transitions
occur at higher rate than transversions \citep{K2P,K3P}. 

It is the purpose of this paper to explore a larger hierarchy of ``RY''  Lie Markov models sensitive to the grouping of nucleotides into purines (R) and pyrimidines (Y).
This hierarchy was derived in \citet{fernandez2014} and totals 37  models capable of distinguishing transitions from transversions. 
To illustrate the various technical issues that arise when using these models,  in section~\ref{sec:5.6b} we explore in detail a relatively simple five-parameter model taken from the hierarchy.
In section~\ref{sec:construction}, we give a construction of the Lie Markov models in a manner that is friendly to non-mathematicians, and discuss how the hierarchy of purine/pyrimidine models can be extended to include models distinguishing different DNA substitution pairs.
In section~\ref{sec:Likelihood-Testing}, we test the Lie Markov models for biological plausibility on a range of real data sets comparing directly to known, popular models (particularly GTR and HKY \citep{HKY}). 
In section~\ref{sec-model-structure}, we examine the nesting relationships of the models within the hierarchy.
In section~\ref{sec:Alternative-parameterizations}, we  present some alternative parameterizations of the model hierarchy, and in section~\ref{sec:stochasticity} explore issues around the stochasticity of ``average'' rate matrices.

\subsection*{Acknowledgements}
We would like to thank Barbara Holland for constructive criticism of manuscript. 
This research was conducted with support from ARC Future Fellowship FT100100031, JGS was supported by Discovery Early Career Fellowship DE130100423, and JFS was supported by the Spanish Government MTM2012-38122-C03-01, and Generalitat de Catalunya, 2014SGR 634. 

\section{Example Lie Markov model: RY5.6b\label{sec:5.6b}}

To motivate the rest of our discussion, we start by presenting one
of the RY Lie Markov models in detail. First we note some notational conventions
and definitions. The column of a rate or Markov matrix indicates the
initial state of the base, and the row the final state, hence rate
matrices have columns which sum to zero (and Markov matrix columns
sum to one). 
The rows and columns are indexed by the DNA bases in
the order A, G, C, T. 
This deviation from standard alphabetical order groups the purines and pyrimidines, making the relations among matrix entries more
apparent. 
The term ``stochastic'' when applied to a rate matrix means that all off-diagonal entries are non-negative, and when applied to a Markov matrix means all entries are non-negative.
We refer to the number of independent parameters in a Lie Markov model as its ``dimension''.

The rate matrices of model RY5.6b can be expressed as

\begin{equation}
\label{eq:5.6b}
Q_{5.6b}=\left(\begin{array}{cccc}
-3a-a_{1}+d+e_{1} & a+a_{1}+d+e_{1} & a+d+e_{1} & a+d+e_{1}\\
a+a_{1}+d-e_{1} & -3a-a_{1}+d-e_{1} & a+d-e_{1} & a+d-e_{1}\\
a-a_{2}-d+e_{2} & a-a_{2}-d+e_{2} & -3a-a_{1}-d+e_{2} & a+a_{1}-d+e_{2}\\
a-a_{2}-d-e_{2} & a-a_{2}-d-e_{2} & a+a_{1}-d-e_{2} & -3a-a_{1}-d-e_{2}
\end{array}\right).
\end{equation}
The ``5'' in the model name indicates that this is a five dimensional model, with parameters $a,a_{2},d,e_{1},e_{2}$. (The choice of parameter labels will be explained in section~\ref{sec:construction}). Note that we can multiply $Q_{5.6b}$ by a scalar and remain in the model. 
If one prefers to think of models as containing only rate matrices of a given scale (e.g. fixed trace) then this is a four dimensional model.
(The trace of a matrix is the sum of elements on the main diagonal, and for a rate matrix acting on a sequence with equal base frequencies, the trace is proportional to the mutation rate.)
Note that the entries of the rate matrix are linear expressions in the parameters; this is a feature of all Lie Markov models, but not of the GTR and related models.

The reader should be alarmed by the appearance of minus signs in the off-diagonal entries of the rate matrix in equation~(\ref{eq:5.6b}).
Unfortunately, there are no simple constraints on the parameters $a,a_{2},d,e_{1},e_{2}$ which restrict to exactly the set of stochastic matrices of this form. 
A reformulation solves this problem and illuminates the model structure significantly: 
\begin{equation}
Q_{5.6b}=\left(\begin{array}{cccc}
* & \alpha+\rho_{A} & \beta+\rho_{A} & \beta+\rho_{A}\\
\alpha+\rho_{G} & * & \beta+\rho_{G} & \beta+\rho_{G}\\
\beta+\rho_{C} & \beta+\rho_{C} & * & \alpha+\rho_{C}\\
\beta+\rho_{T} & \beta+\rho_{T} & \alpha+\rho_{T} & *
\end{array}\right)\label{eq:5.6b-rays}
\end{equation}
where the ``$*$'' stand for the values required for the columns to
sum to zero. 
Now $Q$ is stochastic so long as the parameters are all positive: $\alpha,\,\beta\,,\rho_{A},\,\rho_{G},\,\rho_{C},\,\rho_{T}\geq0$, but the cost of this reformulation is that we are now using six parameters to express a five dimensional model. 
The resulting parameter redundancy is expressed by
\[
Q_{5.6b}(\alpha,\beta,\rho_{A},\rho_{G},\rho_{C},\rho_{T})=Q_{5.6b}(\alpha+\delta,\beta+\delta,\rho_{A}-\delta,\rho_{G}-\delta,\rho_{C}-\delta,\rho_{T}-\delta),
\]
for all choices $\delta$.
The ability to express the model with six non-negative parameters is due to the set of stochastic rate matrices of  this model forming a polyhedral cone having six ``rays'', this being the origin of the ``6'' in the model name. 
Rays are more fully explained in \citet{fernandez2014}.

While all the Lie Markov models can be formulated in this way, most of them acquire redundant parameters -- in some cases \emph{many} redundant parameters -- to ensure stochastic rate matrices.
 In section~\ref{sec:Alternative-parameterizations} we will explore some alternative parameterizations which generate, for computational purposes, the set of stochastic rate matrices of a Lie Markov model with relatively simple parameter constraints and without redundant parameters.

The matrix~(\ref{eq:5.6b-rays}) also reveals that model 5.6b can be thought of as the sum of the Kimura two substitution type (K2ST) model \citep{K2P} (parameters $\alpha,$ $\beta$) and the F81 model \citep{F81} (parameters $\rho_{A}$, $\rho_{G}$, $\rho_{C}$, $\rho_{T}$). 
If we changed the additions in matrix~(\ref{eq:5.6b-rays}) to multiplication, we would have the HKY model \citep{HKY}. 
Most of the Lie Markov models are not so easily related to existing models.

The defining features of the RY Lie Markov models (illustrated here by RY5.6b) are two fold: 
Firstly, the Markov matrices obtained from this model are closed under matrix multiplication (this is what makes the model ``Lie Markov'').
This means that if $M_1$ and $M_2$ are Markov matrices obtained by taking the matrix exponential of two (distinct) rate matrices from the model, then we can expect the product $M_1M_2$ to be obtainable as the matrix exponential of a third RY5.6b rate matrix.
Secondly, the model recognizes the groupings of nucleotides into purines and pyrimidines (this is easily seen by inspection of matrix~(\ref{eq:5.6b-rays})).
The simple idea is that any interchange of nucleotides that preserves the purine/pyrimidine grouping will correspond to a row and column permutation of an RY5.6b rate matrix that will produce another RY5.6b rate matrix.

It is also worth noting that model RY5.6b can have any equilibrium frequencies of bases (under a suitable choice of rate parameters). 
The easiest way to see this is to notice model 5.6b has F81 as a submodel, and F81 can have any equilibrium base frequencies. 
This is not a general property of Lie Markov models --- as noted above, the Jukes-Cantor (JC) and Kimura 3 substitution type (K3ST) models are Lie Markov models, but these have uniform base frequencies at equilibrium. 
The details of the equilibrium base frequencies achievable by the Lie Markov models are given in section~\ref{sec-model-structure}.

\section{Constructing the Lie Markov Models}
\label{sec:construction}

Under a continuous-time formulation with time parameter $t$, a Markov
matrix $M$, whose elements are the probabilities of nucleotide substitutions,
is constructed from a rate matrix $Q$ by matrix exponentiation: 
\[
M=\exp(Qt)=I+Qt+\frac{Q^{2}t^{2}}{2!}+\frac{Q^{3}t^{3}}{3!}+\ldots
\]
Fix a model (e.g. GTR or Kimura's K2ST), and take any two rate matrices $Q_{1}$ and $Q_{2}$ from the model.
Suppose there exists stochastic $\widehat{Q}$ such that $\exp(\widehat{Q}(t_{1}+t_{2}))=\exp(Q_{1}t_{1})\exp(Q_{2}t_{2})$, we would like $\widehat{Q}$ to be in the same model. 
Putting aside the caveat ``if $\widehat{Q}$ exists'' -- in most cases $\widehat{Q}$ will
exist as long as $Q_{1}$ and $Q_{2}$ are not too different -- this would appear a natural condition to ask of a model, especially if one expects some time-inhomogeneity in the DNA substitution process. 

For this property to hold for an given Markov model, \citet{LieMarkov} have shown that is a sufficient condition that the subset of rate matrices
that define the model be 
\begin{lyxlist}{00.00.0000}
\item [{(i)}] closed under addition and scalar multiplication (i.e. the set forms a vector space), 
and 
\item [{(ii)}] closed under matrix \emph{commutator (Lie) brackets},
i.e.. $\left[Q_{1},Q_{2}\right]:=Q_{1}Q_{2}-Q_{2}Q_{1}$ is also in
the space. 
\end{lyxlist}
For the purpose of these conditions we are forced to include non-stochastic rate matrices in the discussion, for example $\left[Q_{1},Q_{2}\right]$
is often not stochastic. 
Together these conditions demand that the model forms a \emph{Lie algebra}. 
Any continuous time Markov model which satisfies these conditions is referred to as a ``Lie Markov model''.

As stated in the introduction, \citet{LieMarkov} derived the set of Lie Markov models that treat each nucleotide on an equal footing. 
\citet{fernandez2014} went further and characterized the ``RY'' Lie Markov models which have a symmetry condition that allow one pairing of DNA bases (canonically the RY pairing: AG and CT) to be treated differently from other pairings.  
We reiterate the essential results here without further discussion as to how they were obtained.

Each RY Lie Markov model has rate matrices which are a linear combination
of basis matrices chosen from a set of 12 (table~\ref{tab:basis}).
Not all subsets of these basis matrices yield a Lie Markov model.
The list of the 37 that do is given in table~\ref{tab:models}.
We adopt a convention that the variable used for the weight of a basis
matrix is the same as the basis matrix name, but in lower-case, e.g.
$e_{1}$ is the weight of $E_{1}$, hence the choice of variable names
in matrix~(\ref{eq:5.6b}).

\begin{table}
\begin{centering}
\begin{tabular}{ccc}
$A=\left(\begin{array}{cccc}
-3 & +1 & +1 & +1\\
+1 & -3 & +1 & +1\\
+1 & +1 & -3 & +1\\
+1 & +1 & +1 & -3
\end{array}\right)$  & $A_{1}=\left(\begin{array}{cccc}
-1 & +1 & 0 & 0\\
+1 & -1 & 0 & 0\\
0 & 0 & -1 & +1\\
0 & 0 & +1 & -1
\end{array}\right)$  & $C=\left(\begin{array}{cccc}
0 & 0 & +1 & -1\\
0 & 0 & -1 & +1\\
-1 & +1 & 0 & 0\\
+1 & -1 & 0 & 0
\end{array}\right)$\tabularnewline
$B=\left(\begin{array}{cccc}
0 & 0 & +1 & -1\\
0 & 0 & -1 & +1\\
+1 & -1 & 0 & 0\\
-1 & +1 & 0 & 0
\end{array}\right)$  & $D_{1}=\left(\begin{array}{cccc}
-1 & +1 & 0 & 0\\
+1 & -1 & 0 & 0\\
0 & 0 & +1 & -1\\
0 & 0 & -1 & +1
\end{array}\right)$  & $D=\left(\begin{array}{cccc}
+1 & +1 & +1 & +1\\
+1 & +1 & +1 & +1\\
-1 & -1 & -1 & -1\\
-1 & -1 & -1 & -1
\end{array}\right)$\tabularnewline
$E_{1}=\left(\begin{array}{cccc}
+1 & +1 & +1 & +1\\
-1 & -1 & -1 & -1\\
0 & 0 & 0 & 0\\
0 & 0 & 0 & 0
\end{array}\right)$  & $F_{1}=\left(\begin{array}{cccc}
+1 & +1 & -1 & -1\\
-1 & -1 & +1 & +1\\
0 & 0 & 0 & 0\\
0 & 0 & 0 & 0
\end{array}\right)$  & $G_{1}=\left(\begin{array}{cccc}
+1 & -1 & \makebox[2\width][c]{0} & \makebox[2\width][c]{0}\\
+1 & -1 & 0 & 0\\
-1 & +1 & 0 & 0\\
-1 & +1 & 0 & 0
\end{array}\right)$\tabularnewline
$E_{2}=\left(\begin{array}{cccc}
0 & 0 & 0 & 0\\
0 & 0 & 0 & 0\\
+1 & +1 & +1 & +1\\
-1 & -1 & -1 & -1
\end{array}\right)$  & $F_{2}=\left(\begin{array}{cccc}
0 & 0 & 0 & 0\\
0 & 0 & 0 & 0\\
+1 & +1 & -1 & -1\\
-1 & -1 & +1 & +1
\end{array}\right)$  & $G_{2}=\left(\begin{array}{cccc}
\makebox[2\width][c]{0} & \makebox[2\width][c]{0} & +1 & -1\\
0 & 0 & +1 & -1\\
0 & 0 & -1 & +1\\
0 & 0 & -1 & +1
\end{array}\right)$\tabularnewline
 & $A_{2}=\left(\begin{array}{cccc}
0 & +2 & -1 & -1\\
+2 & 0 & -1 & -1\\
-1 & -1 & 0 & +2\\
-1 & -1 & +2 & 0
\end{array}\right)$  & \tabularnewline
\end{tabular}
\par\end{centering}

\caption{The rate matrices of RY Lie Markov models are linear combinations of basis matrices.  Each model uses a subset of the twelve matrices listed here. Under some circumstances it is mathematically convenient to replace $A_1$ with the thirteenth matrix, $A_{2}=3A_1-A$.}
\label{tab:basis}
\end{table}

\begin{table}
\resizebox{\textwidth}{!}{ 
\begin{centering}
\begin{tabular}{|cc|cc|}
\hline 
Name  & Basis Matrices  & Name  & Basis Matrices \tabularnewline
\hline 
1.1 (JC) & $A$  & 6.6  &  $A,\, A_1,\, B,\, C,\, D,\, D_{1}$ \tabularnewline
2.2b (K2ST)  & $A,\, A_1$ & 6.7a  &  $A,\, A_1,\, B,\, D,\, E_{1},\,E_{2}$ \tabularnewline
3.3a (K3ST)  & $A,\, A_1,\, B$  & 6.7b  &  $A,\, A_1,\, C,\, D,\, E_{1},\,E_{2}$ \tabularnewline
3.3b  &  $A,\, A_1,\, C$  & 6.8a  &  $A,\, A_1,\, D,\, D_{1},\, E_{1},\,E_{2}$ \tabularnewline
3.3c (TrNef)  &  $A,\, A_1,\, D_{1}$  & 6.8b  &  $A,\, A_1,\, D,\, D_{1},\, G_{1},\,G_{2}$ \tabularnewline
3.4  &  $A,\, A_1,\, D$  & 6.17a  &  $A,\, A_1,\, B,\, D,\, G_{1},\,G_{2}$ \tabularnewline
4.4a (F81)  &  $A,\, D,\, E_{1},\, E_{2}$  & 6.17b  &  $A,\, A_1,\, C,\, D,\, G_{1},\,G_{2}$ \tabularnewline
4.4b  &  $A,\, A_1,\, D,\, D_{1}$  & 8.8  &  $A,\, A_1,\, D,\, D_{1},\, E_{1},\,E_{2},\, F_{1},\,F_{2}$ \tabularnewline
4.5a  &  $A,\, A_1,\, B,\, D$  & 8.10a  &  $A,\, A_1,\, B,\, C,\, D,\, D_{1},\, E_{1},\,E_{2}$ \tabularnewline
4.5b  &  $A,\, A_1,\, C,\, D$  & 8.10b  &  $A,\, A_1,\, B,\, C,\, D,\, D_{1},\, G_{1},\,G_{2}$ \tabularnewline
5.6a  &  $A,\, A_1,\, B,\, C,\, D_{1}$  & 8.16  &  $A,\, A_1,\, D,\, D_{1},\, E_{1},\,E_{2},\, G_{1},\,G_{2}$ \tabularnewline
5.6b  &  $A,\, A_1,\, D,\, E_{1},\,E_{2}$  & 8.17  &  $A,\, A_1,\, B,\, D,\, E_{1},\,E_{2},\, G_{1},\,G_{2}$ \tabularnewline
5.7a  &  $A,\, A_1,\, B,\, E_{1},\,E_{2}$  & 8.18  &  $A,\, A_1,\, B,\, D,\, E_{1},\,E_{2},\, F_{1},\,F_{2}$ \tabularnewline
5.7b  &  $A,\, A_1,\, B,\, F_{1},\,F_{2}$  & 9.20a  &  $A,\, A_1,\, B,\, C,\, D_{1},\, E_{1},\,E_{2},\, F_{1},\,F_{2}$ \tabularnewline
5.7c  &  $A,\, A_1,\, B,\, G_{1},\,G_{2}$  & 9.20b  &  $A,\, A_1,\, B,\, C,\, D_{1},\, F_{1},\,F_{2},\, G_{1},\,G_{2}$ \tabularnewline
5.11a  &  $A,\, A_1,\, D_{1},\, E_{1},\,E_{2}$  & 10.12  &  $A,\, A_1,\, B,\, C,\, D,\, D_{1},\, E_{1},\,E_{2},\, F_{1},\,F_{2}$ \tabularnewline
5.11b  &  $A,\, A_1,\, D_{1},\, F_{1},\,F_{2}$  & 10.34  &  $A,\, A_1,\, B,\, C,\, D,\, D_{1},\, E_{1},\,E_{2},\, G_{1},\,G_{2}$ \tabularnewline
5.11c  &  $A,\, A_1,\, D_{1},\, G_{1},\,G_{2}$  & \multirow{2}{*}{12.12 (GM)}  &  $A,\, A_1,\, B,\, C,\, D,\, D_{1},$ \tabularnewline
5.16  &  $A,\, A_1,\, D,\, G_{1}\,G_{2}$  & & $E_{1},\,E_{2},\, F_{1},\,F_{2},\, G_1,\,G_2$ \tabularnewline
\hline 
\end{tabular}
\par\end{centering}
}
\caption{The RY Lie Markov models. Basis matrix $A_2$ can be substituted for $A_1$ throughout. 
The number before the point indicates the dimension (number of parameters) of the model, the number after the point is the number of rays generated by the model. }
\label{tab:models}
\end{table}

If we take the basis matrices in table~\ref{tab:basis} as having
rows and columns labeled in our canonical order A, G, C, T, then AG
and CT are the distinguished pairings, and we can describe this
as an RY model (puRine/pYrimidine). 
Alternatively if we label the basis matrices in the order A, T, C, G, we distinguish the Watson-Crick pairs AT and CG, which we describe as a WS (Weak/Strong) model.
Finally if we label the basis matrices in order A, C, G, T we distinguish
AC and GT and call these MK (aMino,Keto) models. 
(R, Y, W, S, M and K are the standard IUPAC ambiguity codes for these pairings.)
This allows us to distinguish RY5.6b as model 5.6b with the RY grouping,
whereas model WS5.6b has the same structure but distinguishes AT
and CG (i.e. matrix~(\ref{eq:5.6b}) with row/column ordering A, T, C, G), and similarly MK5.6b is matrix~(\ref{eq:5.6b}) with row/column ordering A, C, G, T. 

If we make statements about (e.g.) the 5.6b model without ``RY'', ``WS'' or ``MK'' prefix, the statement applies equally to RY5.6b, WS5.6b and MK5.6b. 
Additionally, some of the models have full symmetry, meaning there is no distinction between the RY, WS and MK variants. 
These are models 1.1 (Jukes-Cantor), 3.3a (Kimura 3ST), 4.4a (F81), 6.7a (F81+K3ST), 9.20b (doubly stochastic) and 12.12 (general Markov). These models never get a two letter prefix.
Since there are 37 models listed in table~\ref{tab:models}, 31 of which have distinct RY,
WS and MK variants, we have 99 distinct models in total. 
By comparison, the original ModelTest program\citep{modeltest} compares 14 models and jModelTest2\citep{jModelTest2} compares up to 406 models. 
(These counts are before considering rate variation across sites.)

A number of these models have already been studied: 1.1 is the Jukes
Cantor model \citep{JC}, RY2.2b and 3.3a are the Kimura two and three
substitution type models \citep{K2P,K3P} (also known as K2ST, K2P, K80 and K3ST, K3P, K81 models), RY3.3c is the Tamura Nei
model with equal base frequencies \citep{TamuraNei}, 4.4a is the
F81 model \citep{F81}, WS6.6 is the strand symmetric model \citep{yap2004,ssm},
9.20b is the doubly stochastic model  and 12.12 is the general Markov
model \citep{GMM}.

For the purpose of easy comparison to the presentation given in \citet{fernandez2014}, note that we have renamed the basis matrices, added $A_2$ as an alternative to $A_1$, and omitted model 2.2a, which is of no phylogenetic interest as it forbids transversions entirely. 
A table of the basis matrix renaming is in the supplementary material.

\section{Likelihood Testing on Real Data\label{sec:Likelihood-Testing}}

We proceed to investigate how well these models fit real data. We
have taken seven diverse aligned DNA data sets and calculated the
maximum likelihood under each model. The data sets were chosen to
cover a range of DNA types (nuclear, mitochondrial, chloroplast) and
phylogenetic ranges (within a single species to covering a class.)

The data sets are 53 taxa $\times$ 16589 sites human mitochondria
(of which only 202 sites are variable) \citep{human53}, $15\times89436$
(taxa $\times$ sites) angiosperm (+outgroup) chloroplast \citep{acorus15},
$33\times1141$ cormorants and shags, mixed mitochondria and nuclear
\citep{cormorants}, $8\times127026$ \emph{Saccharomyces} (+outgroup)
yeast mostly nuclear plus some mitochondria \citep{rokas}, $11\times2178$
teleost fish nuclear \citep{fish}, $14\times4135$ buttercup (\emph{Rununculus})
chloroplast \citep{buttercup} and $27\times7324$ Ratite (bird order)
mitochondria \citep{ratites}. 

The models tested are the 99 Lie Markov models discussed above (6 fully
symmetric, 31 with RY, WS and MK variants) and, for comparison, the time reversible models of the original ModelTest program \citep{modeltest}. ModelTest uses fourteen models, but 
five of these are also RY Lie Markov models (JC=1.1, K80=RY2.2b, K81=3.3a, TrNef=3.3c, F81=4.4a) so this adds nine models for a total of 108.

Our analysis imitates the procedure used by ModelTest \citep{modeltest}:
(i) A neighbor joining tree is created using the Jukes-Cantor distances;
(ii) The tree is then midpoint rooted (as most RY Lie Markov models are not
time reversible, root location is relevant); (iii) For each model,
we find the maximum likelihood by optimizing model parameters and
branch lengths (but not tree topology) using a hill climbing algorithm (the base distribution at the root is assumed equal to the equilibrium
distribution of the model); (iv) The optimization is performed for
four different models of rate variations across site: single rate,
invariant sites (+I), Gamma rate distribution (+$\Gamma$, with 8
rate classes) and both invariant sites and Gamma distribution (+I+$\Gamma$);
(v) Finally, we apply the Bayesian Information Criterion (BIC) \citep{BIC}\emph{
}correction to penalize models with more parameters (Table~\ref{tab:BIC}).

\begin{table}
\resizebox{\textwidth}{!}{
\begin{tabular}{|l|r|r|r|r|r|r|r|}
\hline 
Clade:  & \multicolumn{1}{c|}{Human} & \multicolumn{1}{c|}{Angiosperms} & \multicolumn{1}{c|}{Cormorants} & \multicolumn{1}{c|}{Yeast} & \multicolumn{1}{c|}{Teleost Fish} & \multicolumn{1}{c|}{Buttercups} & \multicolumn{1}{c|}{Ratites}\tabularnewline
Approx range:  & \multicolumn{1}{c|}{Species} & \multicolumn{1}{c|}{Class} & \multicolumn{1}{c|}{Family} & \multicolumn{1}{c|}{Genus} & \multicolumn{1}{c|}{mult. orders } & \multicolumn{1}{c|}{Genus} & \multicolumn{1}{c|}{Order}\tabularnewline
DNA type  & \multicolumn{1}{c|}{mitoch} & \multicolumn{1}{c|}{chlorop} & \multicolumn{1}{c|}{mito/nuc} & \multicolumn{1}{c|}{mostly nuc} & \multicolumn{1}{c|}{nuclear} & \multicolumn{1}{c|}{chlorop} & \multicolumn{1}{c|}{mitoch}\tabularnewline
taxa$\times$sites  & \multicolumn{1}{c|}{$53\times16589$} & \multicolumn{1}{c|}{$15\times89436$} & \multicolumn{1}{c|}{$33\times1141$} & \multicolumn{1}{c|}{$8\times127026$} & \multicolumn{1}{c|}{$11\times2178$} & \multicolumn{1}{c|}{$14\times4135$} & \multicolumn{1}{c|}{$27\times7324$}\tabularnewline
Site rate model  & \multicolumn{1}{c|}{$+\Gamma+I$} & \multicolumn{1}{c|}{$+\Gamma$} & \multicolumn{1}{c|}{$+\Gamma+I$} & \multicolumn{1}{c|}{$+\Gamma+I$} & \multicolumn{1}{c|}{$+\Gamma$} & \multicolumn{1}{c|}{$+I$} & \multicolumn{1}{c|}{$+\Gamma+I$}\tabularnewline
\hline 
$1^{\rm st}$ & TrN     & MK10.34 & HKY     & 12.12   & RY5.11b & WS4.4b & RY8.16 \\
\hline
$2^{\rm nd}$ & HKY     & RY8.18  & TrN     & GTR     & RY3.3c  & WS3.4  & RY10.34\\
$\Delta BIC$ & 8.90    & 16.17   & 6.47    & 79.77   & 0.08    & 0.02   & 7.87\\
\hline
$3^{\rm rd}$ & TIM     & 12.12   & K81uf   & RY10.12 & RY2.2b  & WS4.5a & TVM\\
$\Delta BIC$ & 9.68    & 17.09   & 6.78    & 911.96  & 3.45    & 5.12   & 8.65\\
\hline
$4^{\rm th}$ & RY8.8   & MK8.17  & RY8.8   & RY8.8   & RY5.7b  & WS4.5b & 12.12\\
$\Delta BIC$ & 13.53   & 26.94   & 10.82   & 946.46  & 6.45    & 5.99   & 14.05\\
\hline
$5^{\rm th}$ & RY8.18  & WS8.10a & TIM     & RY9.20a & TIMef   & MK5.7a & GTR\\
$\Delta BIC$ & 15.44   & 27.50   & 13.29   & 1156.55 & 6.69    & 10.33  & 15.19\\
\hline
$6^{\rm th}$ & K81uf   & WS10.12 & RY8.18  & WS10.12 & RY4.4b  & RY5.7a & WS10.12\\
$\Delta BIC$ & 18.58   & 28.37   & 16.18   & 1450.58 & 7.54    & 10.74  & 18.57\\
\hline
$7^{\rm th}$ & GTR     & RY10.12 & MK8.10a & TVM     & RY3.4   & WS6.8a & RY8.17\\
$\Delta BIC$ & 21.29   & 29.87   & 17.14   & 1518.73 & 8.52    & 11.49  & 30.55\\
\hline
$8^{\rm th}$ & TVM     & WS10.34 & TVM     & TIM     & SYM     & WS5.6b & WS8.10a\\
$\Delta BIC$ & 29.92   & 36.24   & 19.32   & 1613.45 & 9.51    & 12.36  & 34.76\\
\hline
$9^{\rm th}$ & RY10.12 & RY9.20a & RY10.12 & TrN     & RY5.11a & WS6.6  & MK10.12\\
$\Delta BIC$ & 30.50   & 89.65   & 20.24   & 1640.46 & 9.57    & 13.05  & 42.28\\
\hline
$10^{\rm th}$& MK10.34 & WS6.6   & WS8.17  & MK10.34 & 3.3a    & K81uf  & WS10.34\\
$\Delta BIC$ & 31.09   &108.59   & 21.27   & 1663.04 & 10.05   & 14.37  & 45.58\\
\hline
\end{tabular}}
  
\caption{The top 10 models for each data set, by Bayesian Information Criterion (BIC). $\Delta BIC$ is how much worse this model scores than the optimal model ($1^{\rm st}$). A complete table of BIC scores is available in the supplementary material. \label{tab:BIC}}
\end{table}

In Table~\ref{tab:BIC} we present BIC scores for each model under the optimal
rate variation across sites model. 
(Scores for non-optimal rate variation models are in the supplementary material spreadsheet.)
For each data set, the models were ranked by BIC and, for each model,
these rankings are summarized in Table~\ref{tab:BIC ranks}. 
\begin{table}
\resizebox{\textwidth}{!}{

\begin{tabular}{|c|ccc|}
\hline 
Model  & Mean  & Best  & EBF\tabularnewline
 & rank  & rank  & DF\tabularnewline
\hline 
RY8.18  & 14.4 & 2  & 3 \\
*TVM     & 15.1 & 3  & 3 \\
*K81uf   & 16.9 & 3  & 3 \\
*TIM     & 16.9 & 3  & 3 \\
*GTR     & 17.6 & 2  & 3 \\
RY8.8   & 18.0 & 4  & 3 \\
RY10.12 & 19.6 & 3  & 3 \\
*HKY     & 20.0 & 1  & 3 \\
*TrN     & 20.0 & 1  & 3 \\
MK10.34 & 21.1 & 1  & 3 \\
12.12   & 21.7 & 1  & 3 \\
WS8.10a & 21.9 & 5  & 3 \\
WS10.34 & 22.3 & 8  & 3 \\
RY9.20a & 23.0 & 5  & 2 \\
MK8.17  & 23.1 & 4  & 3 \\
MK8.10a & 24.3 & 7  & 3 \\
WS10.12 & 25.6 & 6  & 3 \\
6.7a    & 25.9 & 13 & 3 \\
WS8.17  & 26.4 & 10 & 3 \\
RY6.8a  & 27.0 & 13 & 3 \\
RY8.10a & 27.1 & 11 & 3 \\
MK10.12 & 27.3 & 9  & 3 \\
RY10.34 & 28.7 & 2  & 3 \\
RY8.17  & 28.7 & 7  & 3 \\
RY8.16  & 29.6 & 1  & 3 \\
RY5.11a & 29.9 & 9  & 2 \\
MK8.18  & 30.0 & 16 & 3 \\
\hline 
\end{tabular}~~~~%
\begin{tabular}{|c|ccc|}
\hline 
Model  & Mean  & Best  & EBF\tabularnewline
 & rank  & rank  & DF\tabularnewline
\hline 
RY5.7a  & 30.0 & 6  & 2 \\
RY5.6b  & 30.3 & 21 & 3 \\
MK5.7a  & 30.6 & 5  & 2 \\
WS8.18  & 32.9 & 24 & 3 \\
RY6.7b  & 34.0 & 22 & 3 \\
MK9.20a & 36.1 & 19 & 2 \\
WS6.6   & 37.1 & 9  & 1 \\
WS4.5a  & 37.7 & 3  & 1 \\
WS6.17a & 40.0 & 15 & 1 \\
WS8.10b & 40.3 & 10 & 1 \\
WS5.7a  & 43.6 & 23 & 2 \\
*SYM     & 46.0 & 8  & 0 \\
MK6.6   & 47.4 & 18 & 1 \\
MK8.10b & 47.6 & 17 & 1 \\
*TVMef   & 48.1 & 13 & 0 \\
WS9.20a & 48.3 & 28 & 2 \\
MK4.5a  & 49.6 & 25 & 1 \\
MK6.17a & 50.1 & 32 & 1 \\
RY4.5a  & 50.6 & 18 & 1 \\
RY6.17a & 51.9 & 34 & 1 \\
RY6.6   & 52.4 & 33 & 1 \\
RY6.8b  & 52.7 & 34 & 1 \\
RY8.10b & 52.7 & 33 & 1 \\
RY4.4b  & 53.1 & 6  & 1 \\
*TIMef   & 54.6 & 5  & 0 \\
RY5.11b & 54.7 & 1  & 0 \\
MK5.6a  & 56.0 & 19 & 0 \\
\hline 
\end{tabular}~~~~%
\begin{tabular}{|c|ccc|}
\hline 
Model  & Mean  & Best  & EBF\tabularnewline
 & rank  & rank  & DF\tabularnewline
\hline 
RY5.7b  & 56.0 & 4  & 0 \\
RY5.16  & 56.4 & 39 & 1 \\
9.20b   & 56.6 & 39 & 0 \\
RY5.11c & 56.7 & 15 & 0 \\
RY3.4   & 56.7 & 7  & 1 \\
WS5.6a  & 56.7 & 23 & 0 \\
RY5.6a  & 58.3 & 14 & 0 \\
3.3a    & 59.1 & 10 & 0 \\
RY4.5b  & 59.4 & 22 & 1 \\
RY5.7c  & 59.7 & 43 & 0 \\
RY6.17b & 60.0 & 35 & 1 \\
MK5.7c  & 60.1 & 17 & 0 \\
MK5.7b  & 60.6 & 37 & 0 \\
RY3.3c  & 60.6 & 2  & 0 \\
WS5.7c  & 60.7 & 16 & 0 \\
WS5.7b  & 63.7 & 40 & 0 \\
RY2.2b  & 64.1 & 3  & 0 \\
WS6.7b  & 66.6 & 14 & 3 \\
RY3.3b  & 66.7 & 11 & 0 \\
WS6.8a  & 66.9 & 7  & 3 \\
WS8.8   & 67.1 & 22 & 3 \\
WS8.16  & 68.0 & 23 & 3 \\
WS5.6b  & 68.6 & 8  & 3 \\
MK6.7b  & 71.1 & 43 & 3 \\
MK6.8a  & 71.1 & 40 & 3 \\
MK8.8   & 71.3 & 39 & 3 \\
MK8.16  & 71.4 & 38 & 3 \\
\hline 
\end{tabular}~~~~%
\begin{tabular}{|c|ccc|}
\hline 
Model  & Mean  & Best  & EBF\tabularnewline
 & rank  & rank  & DF\tabularnewline
\hline 
MK5.6b  & 74.1 & 34 & 3 \\
WS4.5b  & 74.4 & 4  & 1 \\
WS6.17b & 76.0 & 16 & 1 \\
WS4.4b  & 76.9 & 1  & 1 \\
WS3.4   & 78.0 & 2  & 1 \\
WS6.8b  & 78.4 & 12 & 1 \\
WS5.16  & 79.0 & 11 & 1 \\
WS5.11a & 80.3 & 50 & 2 \\
MK5.11a & 80.9 & 37 & 2 \\
MK4.5b  & 87.1 & 69 & 1 \\
MK4.4b  & 88.3 & 64 & 1 \\
WS3.3b  & 88.3 & 59 & 0 \\
4.4a    & 88.9 & 44 & 3 \\
MK6.17b & 89.6 & 71 & 1 \\
MK6.8b  & 90.0 & 67 & 1 \\
WS2.2b  & 90.3 & 57 & 0 \\
WS5.11b & 90.6 & 65 & 0 \\
WS3.3c  & 90.7 & 56 & 0 \\
MK3.4   & 91.3 & 68 & 1 \\
WS5.11c & 92.6 & 67 & 0 \\
MK5.16  & 93.0 & 70 & 1 \\
MK3.3c  & 94.7 & 75 & 0 \\
MK3.3b  & 95.0 & 77 & 0 \\
MK2.2b  & 95.7 & 74 & 0 \\
MK5.11b & 95.9 & 85 & 0 \\
MK5.11c & 97.1 & 78 & 0 \\
JC      & 104.3& 83 & 0 \\
\hline 
\end{tabular}

}

\caption{Summary of rankings of models under BIC for the 7 data sets. Models marked ``*'' are time reversible, non-Lie Markov models. EBF DF
= Equilibrium frequency degrees of freedom (section~\ref{sec-model-structure}).
The best ranked models have high EBF DF.  \label{tab:BIC ranks}}
\end{table}

The best overall ranking goes to the 8 dimensional Lie Markov model RY8.18. 
This model is comparable in complexity to GTR, which has 9 dimensions as we count them.
Although time reversible models ranked first in just two of the seven data sets (table~\ref{tab:BIC}), they rank well overall, holding six spots in the top ten sorted by mean rank (table~\ref{tab:BIC ranks}). 
The top-ranking Lie Markov models were RY8.18 and RY8.8. 
We expect the WS and MK models to do poorly since they do not recognize the established
biological preference for transitions over transversions. 
They do indeed dominate the bottom of the table, however the top of the table shows only slight preference for RY over MK or WS models, with MK10.34 taking tenth position (fourth best of the Lie Markov models). 
We caution against reading too much into these rankings, due to the small size of the sample.

Some models score poorly overall, but score well for one data set. The top ranked models for the fish data set are RY5.11b and RY3.3c  (mean ranks 54.7 and 60.57). The top ranked models for the buttercup data set are WS4.4b and WS3.4 (mean ranks 76.8 and 78.0). We will have more to say on the buttercup results later in this section, and the fish data set in section~\ref{sec:ebf}.

The Corrected Akaike Information Criterion (AICc) \citep{AIC,AICc} penalizes extra parameters much less than the BIC. 
An analysis using AICc in place of BIC is given in the supplementary material. 
Under AICc ranking, the top four models are RY10.12, 12.12 (general Markov model), GTR then RY8.18.

Despite model RY5.6b's structural similarity to HKY (section~\ref{sec:5.6b}),
it does not perform well in comparison to HKY. Model 6.7a (the sum
of K3ST and F81) ranks better, but still well below HKY. 

Model RY8.8 performed well and is of particular interest since it is the smallest Lie Markov model that contains all Markov matrices obtainable by multiplying different HKY Markov matrices (the curious reader will be interested to learn the corresponding closure of GTR is the General Markov model, 12.12).
The RY8.8 rate matrix can be parameterized as 
\[
Q_{8.8}=\left(\begin{array}{cccc}
* & a & e & e\\
b & * & f & f\\
g & g & * & c\\
h & h & d & *
\end{array}\right).
\]

The buttercup data set produced results markedly different from the
rest, highly ranking WS models with few parameters, and ranking RY
models poorly in general. The top four models are all submodels of WS6.6, the strand symmetric model \citep{ssm}. It appears that the
assumptions behind the strand symmetric model, and the WS models generally,
hold for these chloroplast sequences (which are largely intergenic
spacers \citep{buttercup}).

In conclusion, we see that for a given data set, we can generally
find a Lie Markov model which outscores a time reversible model, although time reversible models perform well in comparison to the full set of Lie Markov models.
Model RY8.8 stands out as one of the best performing while also having
theoretical justification as the closure of the HKY model. Unexpectedly,
models with non-standard base pairings (WS and MK) can also score
well for particular models (MK10.34, WS8.10a) or particular data sets
(buttercups).

\section{The Structure of RY Models}
\label{sec-model-structure}

\subsection{Nesting of the models}

\begin{figure}[]
\includegraphics[width=\textwidth]{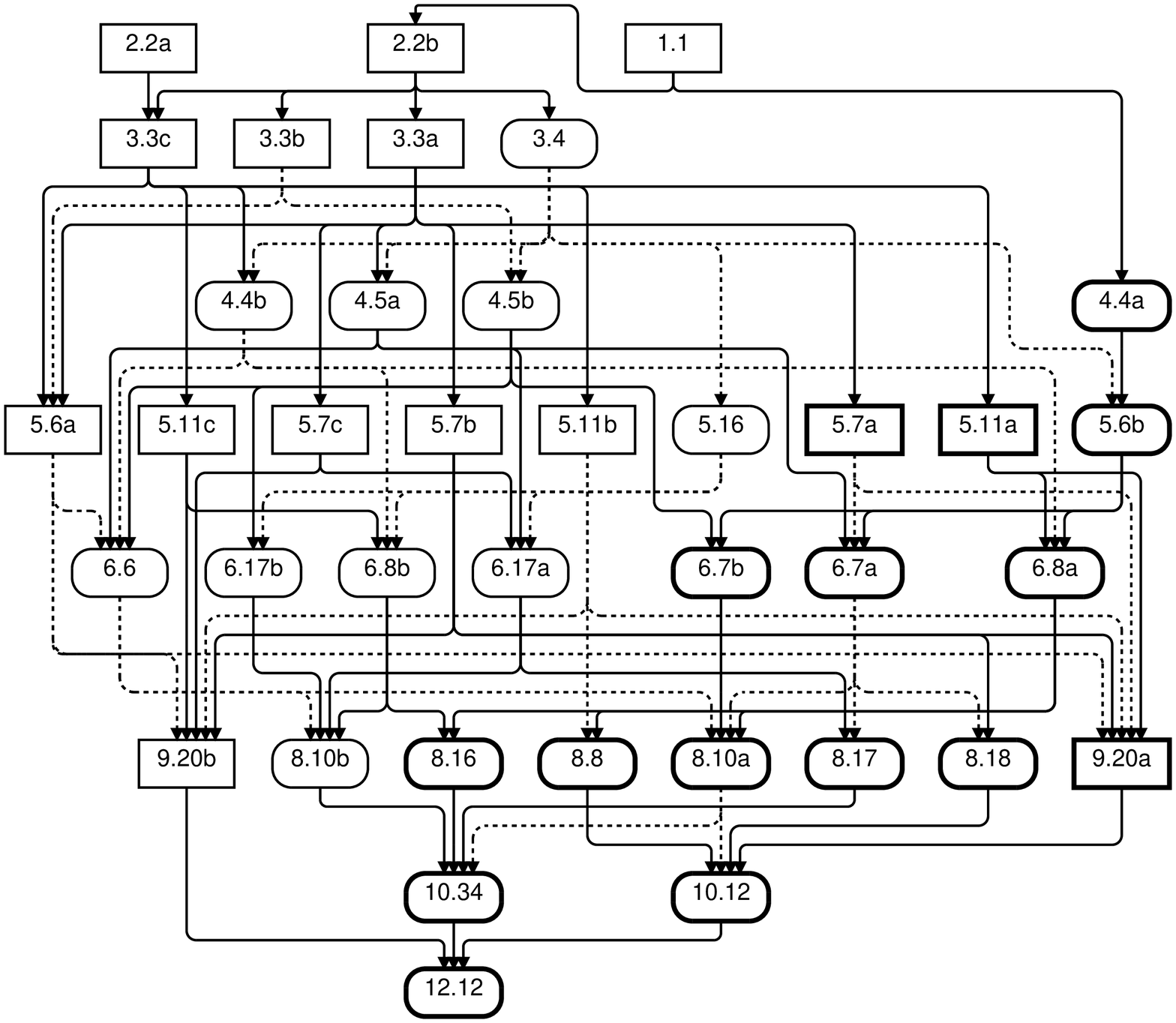}
\protect\vspace{0.25cm}
\begin{center}
\includegraphics[width=12cm]{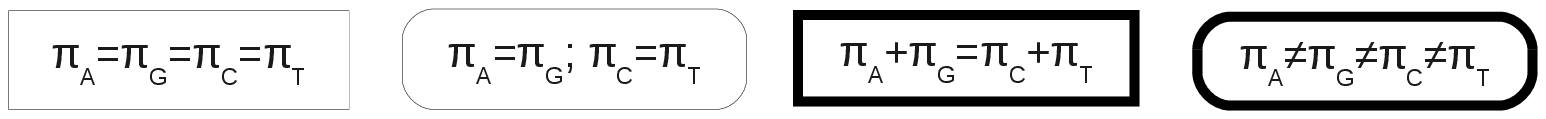}
\end{center}

\caption{Nesting relationships of the RY Lie Markov models. Box shape and line weight indicates the degrees of freedom in equilibrium base frequencies (see text). Solid or dotted connecting lines are to reduce visual confusion and have no additional significance. \label{fig:Nesting}}
\end{figure}

When all the rate matrices in model A also occur in model B, we say
model A is nested within model B --- i.e. by adding constraints to
model B we can create model A. We may wish to know these relationships
so that we can justify using a likelihood ratio test, or to use the
optimal solution for model A as an initial solution for optimizing
model B, or for an MCMC analysis which allows switching between related
models. The nesting relationships of the RY Lie Markov models can easily be derived from the basis matrix specifications of the models, given in Table~\ref{tab:models}.
The hierarchy of nestings is shown in figure~\ref{fig:Nesting}.

Model 6.7a is the F81+K3ST model. This model has full symmetry, and
so is simultaneously in the RY, WS and MK model families.  This means
that in some cases low parameter models in one model family are nested
within high parameter models of another, e.g. RY5.7a, being nested
in 6.7a is (by transitivity) also nested in WS8.10a.

A model is ``doubly stochastic''  if the rows of its rate matrices
always sum to zero (in addition to the columns sum to zero condition
required of a rate matrix). The most general model with this property
is the ``doubly stochastic model'', which is model 9.20b in our
hierarchy. All models nested within 9.20b also have the doubly stochastic
property, e.g. 3.3a (K3ST) and 1.1 (JC). 

\subsection{Equilibrium base frequencies}\label{sec:ebf}

An important property of a model is the range of equilibrium base
frequencies (EBF) it can produce. If the base frequencies in the data
differ greatly from the EBF of the model, a poor likelihood score
is inevitable. The EBF of a given rate matrix is its principal right-eigenvector,
which will have eigenvalue zero (as a consequence of the columns-sum-to-zero
constraint). The same applies for a Markov matrix, except the eigenvalue
will be one.

The doubly stochastic property implies flat EBF, as $(\nicefrac{1}{4},\nicefrac{1}{4},\nicefrac{1}{4},\nicefrac{1}{4})$
is an eigenvector (eigenvalue zero) of any doubly stochastic Markov
matrix, with eigenvalue 1, and hence the EBF for 9.20b has zero degrees
of freedom, with EBF $\pi_{A}=\pi_{G}=\pi_{C}=\pi_{T}=\nicefrac{1}{4}$.
Nine of the basis matrices (table~\ref{tab:basis}) have this doubly stochastic property, those nine being $A,A_1,B,C,D_1,F_1,F_2,G_1\mbox{ and }G_2$, which are also basis matrices of 9.20b, the most general doubly stochastic model. Any model whose basis matrices come from this set will also be doubly stochastic and so have flat equilibrium base frequencies. These models (the submodels of 9.20b) are 1.1, 2.2a, 2.2b, 3.3a, 3.3b, 5.6a, 5.7b, 5.7c, 5.11b, 5.11c. 

The remaining three basis matrices are $D$, $E_1$ and $E_2$. Each matrix adds one degree of freedom to the EBF distribution. The simplest model to contain all three is 4.4a, the F81 model\citep{F81}. This model has the maximum of three degrees of freedom in its EBF since $\pi_{A}+\pi_{G}+\pi_{C}+\pi_{T}=1$. 
Supermodels of 4.4a also have full EBF freedom, being 5.6b, 6.7a, 6.7b, 6.8a, 8.8, 8.10a, 8.16, 8.17, 8.18, 10.12, 10.34 and 12.12.

Models which contain $D$ but not $E_1$ and $E_2$ have $\pi_{A}=\pi_{G}\neq\pi_{C}=\pi_{T}$ (one degree of freedom). 
These models are 3.4, 4.4b, 4.5a, 4.5b, 5.16, 6.6, 6.8b, 6.17a, 6.17b, and 8.10b. 

Any model with $E_1$ or $E_2$ has both, and the models containing these two but not $D$ are 5.7a, 5.11a and 9.20a. The EBF of these models have two degrees of freedom, $\pi_{A}+\pi_{G}=\pi_{C}+\pi_{G}=\nicefrac{1}{2}$.

These degrees of freedom are indicated in figure~\ref{fig:Nesting}. Table~\ref{tab:BIC ranks} demonstrates that models with many EBF degrees of freedom generally outperform those with few degrees of freedom. 
We now can understand the unusual choice of models for the fish data set: the top three models (RY5.11b, RY3.3c and RY2.2b) all have zero EBF degrees of freedom. Because this data set is unusual in having close to flat base frequencies (24.4\% A, 25.2\% G, 23.1\% C, 27.4\% T) it is able to accept these models where the other data sets strongly reject them.

In contrast to GTR, the relationship between EBF and model parameters
for Lie Markov models is often not simple. For example, for model
RY5.6b the EBF are
\[
(\pi_{A},\pi_{G},\pi_{C},\pi_{T})=(\frac{1}{4},\frac{1}{4},\frac{1}{4},\frac{1}{4})+\frac{1}{4p}(q+2e_{1},q-2e_{1},-q+2e_{2},-q-2e_{2})
\]
where $p=2a+a_{1}$ and $q=2d+\frac{a_{1}d}{a}$.

Only a few of the Lie Markov models presented here are time reversible, namely 1.1, 2.2a,
2.2b, 3.3a, 3.3c, 3.4, 4.4a and 4.4b. In the context of a time inhomogeneous mutation process, we expect base frequencies to be out of equilibrium, so a time reversible analysis is inappropriate in any case. In this circumstance, there is no advantage to a time reversible model, so we do not regard the non-reversibility of our models as a major drawback. 
Time reversibility is a computational convenience, not a law of nature.

In passing, we also suggest to the maintainers of jModelTest2 that they consider adding more flexibility to the equilibrium base frequency distributions, as currently they allow only 0 or 3 degrees of freedom.

\section{Parameterizations }
\label{sec:Alternative-parameterizations}

In section~\ref{sec:5.6b} we briefly alluded to the problem of generating rate matrices which are stochastic, i.e. all off diagonal elements are non-negative. We require parameterizations of the Lie Markov models for which (1) simple bounds on the parameters (i.e. not dependent on the values of other parameters) restrict the resulting rate matrices to be stochastic; (2) all stochastic rate matrices in the model can be generated from parameters within the bounds; (3) that slightly different rate matrices can always be specified by slightly different parameters (i.e. the inverse transformation of rate-matrix to parameters is continuous). 

These conditions allow us to conduct likelihood optimizations by hill-climbing. The simple bounds give us a well defined region of parameter space to search. Condition (2) ensures that all legitimate solutions lie within the space to be searched. Condition (3) ensures the hill climb does not get blocked by a parameterization boundary. We will now derive such a parameterization.

A DNA rate matrix is defined by its 12 off-diagonal elements, so DNA rate matrices lie within a 12 dimensional space. This portion of this space which is stochastic can be equated to the general Markov model, and less general models are subsets of it, generally of lower dimension. 

Consider an $n$ dimensional linear (as defined in section~\ref{sec:5.6b}) model, and take the $n$ basis matrix weights as the coordinates of its space. The stochasticity constraint gives us (up to) twelve linear inequalities, each expressing the non-negativity of a given matrix element. The region of this space which is stochastic is therefore the intersection of the half-spaces defined by these constraints, and we further note that the boundaries of these half spaces all pass through the origin. (If all basis matrix weights are zero, all off-diagonal elements are zero.) These facts are enough to establish that the region of stochasticity is a geometric entity known as a convex polyhedral cone. 

In the context of this section, it simplifies matters to take $A_2$ as a basis matrix in place of $A_1$ (table~\ref{tab:basis}). Then, all matrices $B_i\neq A$ from  table~\ref{tab:basis} are orthogonal to $A$, and span the space of rate matrices with trace 0. 
In particular, the scale (trace) of the rate matrix is determined only by $a$, the weight of $A$, and that for fixed $a$, none of the other weights can go to infinity without violating stochasticity. 
It follows that the set of rate matrices with a fixed trace defines a bounded set.

\begin{figure}[]
\begin{center}
\includegraphics[width=12cm]{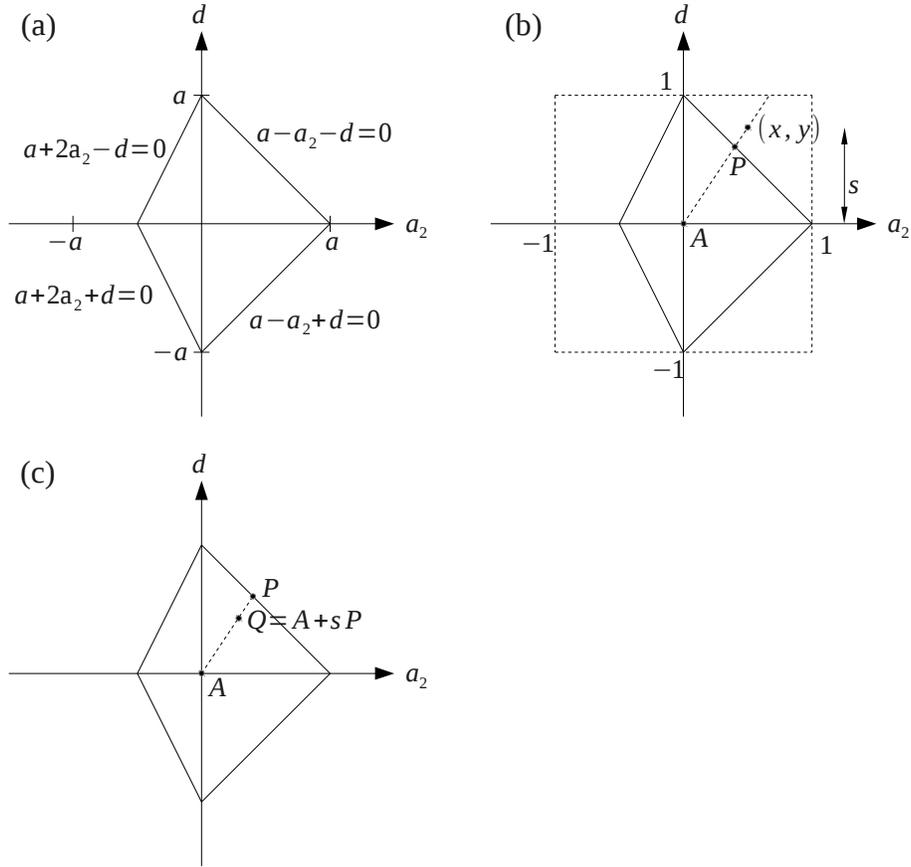}
\end{center}
\caption{A parameterization of model 3.4 which is restricted to only the stochastic rate matrices. (a) The region of stochasticity for model 3.4 with fixed $a$. (b) Without loss of generality, we take $a=1$. Given $(x,y)$ in $[-1,1]$ defines point (representing a matrix) $P$ on the edge of the region of stocasticity, and $s=\max(|x|,|y|)$ a measure of how far $(x,y)$ is from the origin, the Jukes-Cantor matrix $A$. (c) ($x,y$) have defined a stochastic rate matrix $Q(x,y)=A+s P$.\label{fig:3.4}}
\end{figure}

For example, model 3.4 has rate matrix
\[
Q_{3.4}=\left(\begin{array}{cccc}
*        & a+2a_2+d & a-a_2+d  & a-a_2+d\\
a+2a_2+d & *        & a-a_2+d  & a-a+2+d\\
a-a_2-d  & a-a+2-d  & *        & a+2a_2-d\\
a-a_2-d  & a-a_2-d  & a+2a_2-d & *
\end{array}\right),
\]
so the stochasticity constraints can be expressed as 
\begin{equation}
  \begin{split} 
    a+2a_2+d &\geq 0,\\
    a+2a_2-d &\geq 0,\\
    a-a_2+d  &\geq 0,\\
    a-a_2-d  &\geq 0. \label{eq:3.4}
  \end{split}
\end{equation}
This is shown graphically in figure~\ref{fig:3.4}(a). 

We refer to our preferred parameterization of the RY Lie Markov models as the Cartesian parameterization (it is illustrated for model 3.4 in figure~\ref{fig:3.4}(b) and~(c)). 
From a choice of parameters, this parameterization will produce a stochastic rate matrix $Q$ within the model, and with some given trace.
In general, we are given an $n$ dimensional RY Lie Markov model, having basis matrices $A, B_1,\ldots B_{n-1}$ (where the $B_i$ stand for non-$A$ basis matrices from table~\ref{tab:basis} as above). 
Next, we proceed to describe the parameterization in three steps. 
\begin{enumerate}
 \item Generate a (non-trivial) matrix
\begin{eqnarray*}
P'=\sum_{i}b_i B_i,  \mbox{ where all }b_i\in [-1,1].
\end{eqnarray*}
 The weights $(b_1,\ldots,b_{n-1})$ are taken as the parameters. 

\item Define the ``perturbation'' matrix $P$ by
\[
P=\frac{1}{-\min(P')}P'
\]
where $\min P'$ is the minimum off-diagonal element in $P'$. Note all the $B_i$ have off-diagonal elements summing to zero, therefore $P'$ will always contain a negative off-diagonal element (unless it is zero.) Therefore $-\min(P')\geq 0$. 
\item Now we find the ``saturation'' value by
\[
s=\max_i |b_i|
\]
and finally our rate matrix is
\[
Q=A+s P=A-\frac{s}{\min(P')} P'.
\]
\end{enumerate}
If $s=1$, $Q$ will be on the boundary of stochasticity, having (at least) one off-diagonal element equal to zero, as $A$ has all off-diagonal elements equal one.

The map $[-1,1]^{n-1}\mapsto Q$ defined as above is one-to-one, and parameterizes uniformly the section of the stochastic cone with trace $-12$ taking as parameter space the hypercube $[1,1]^{n-1}$.
Should a different fixed scale be desired, we can multiply by a constant. Should we wish the scale of $Q$ to be variable, we can add a scale parameter. 

The essence of this method is that the ratios of the $b_i$ define the direction in which we will deviate from the Jukes-Cantor matrix $A$, and the overall scale of the $b_i$ sets how far we travel from Jukes-Cantor towards the boundary of stochasticity. We can also think of it geometrically, as using the $b_i$ to form a hypercube enclosing the hyperpolyhedron which is the region of stochasticity, and then shrink-wrapping the hypercube around the hyperpolyhedron. While this parameterization gives $Q$ as a continuous function of the $b_i$, it is not a smooth function, and so may not work well with hill climbing methods which calculate partial derivatives. 

We will briefly describe three alternative parameterizations which we explored prior to settling on the Cartesian parameterization described above. Given the stochasticity inequalities (e.g. equations~(\ref{eq:3.4}) for model 3.4) we can progressively eliminate variables by Fourier-Motzkin elimination \citep{Motzkin}. This gives us a parameterization where, having used $x_1,\ldots,x_k$ to set the weights of $B_1,\ldots,B_k$, we know the allowable range of weights for $B_{k+1}$ which will keep stochasticity, and we linearly transform $x_{k+1}$ appropriately. The disadvantage of this parameterization is that we need extra computer code specific to each model to implement the Fourier-Motzkin derived transformation. The {\em Mathematica} file in the supplementary material derives Fourier-Motzkin transformations for each of the models.

The Cartesian parameterization uses the ratios of $n-1$ parameters to determine a direction and the scale of the parameters to determine a distance. We can separate these roles and use $n-2$ parameters to specify a direction and supply the ``saturation'' directly as the $n-1^{\rm th}$ parameter, i.e. we use polar coordinates in the space of matrices with zero trace. In the shrink-wrap analogy described above, this corresponds to shrink-wrapping a hypersphere rather than hypercube. The weakness of this method is that the inverse transformation is non-continuous: $Q$ matrices which are close to each other may not have parameters which are close to each other, due to an angle wrapping from $2\pi$ to zero. 
We tested an extension where angles were unbounded and the radius parameter was in $[-1,1]$ instead of $[0,1]$ (which means the parameter to rate matrix mapping is no longer 1:1.) This helped, but optimization still often failed to find the best likelihood.

Finally we can form a rate matrix as a sum of non-negatively weighted ray matrices. While this is simple to code and gives a continuous and smooth function, most models have more rays than dimensions, so this requires too many parameters.

\section{Stochasticity}
\label{sec:stochasticity}
Multiplicative closure can be tested by taking stochastic rate matrices $Q_{1}$ and $Q_{2}$ from a model and calculating $\widehat{Q}=\log(\exp(Q_{1})\exp(Q_{2}))$~(where ``log'' is the matrix logarithm). 
The desired result is that $\widehat{Q}$ be stochastic and in the model. 

There are three possible failure modes: (i.) the matrix logarithm can produce complex values, so $\widehat{Q}$ may be complex and hence not stochastic; (ii.) $\widehat{Q}$ may be real but not stochastic; or (iii.) $\widehat{Q}$ may not be in the model. 
In the Markov chain literature, the property of $\widehat{Q}$ being stochastic is called ``embeddability'', and is is discussed at length in the context of phylogenetics and
time inhomogeneous DNA models by \citet{Embedding}. 
General Lie theory tells us that the last of these failure modes should not be a possibility for a
Lie Markov model, however we included this possibility in what follows as a sanity check.

We made a preliminary Monte Carlo investigation to get some feeling for how often these failures occur. 
For each model, we repeatedly generate two random rate matrices $Q_{1}$ and $Q_{2}$ within the model and having predetermined trace, and calculate $\widehat{Q}$.
We determine whether this $\widehat{Q}$ is stochastic, real, and in the model. 
As the traces of $Q_{1}$ and $Q_{2}$ get larger, the chances of non-stochastic (or non-real) $\widehat{Q}$ grows. 
In table~\ref{tab:nonstochastic}, we show the level of saturation before about 5\% of random products give a non-stochastic (or non-real) $\widehat{Q}$. 
(1 expected substitution per site corresponds to a trace of -4.) 
By this measure, the worst performing model was 10.12, which achieved this 5\% non-embeddability threshold with trace about $-3.3$. 
We observed no instances of $\widehat{Q}$ not being in the model, even when $\widehat{Q}$ is complex. 
The procedure for randomly selecting rate matrices from within a model is described in the supplementary material, and the calculations are carried out in the supplementary material \emph{Mathematica} notebook.

\begin{table}
\hfill{}%
\begin{tabular}{|c|c|}
\hline 
1 substitution/site & 5.6a, 6.6, 6.8a, 6.8b, 8.8, 8.10a, 8.10b, 8.16, 8.17, 8.18, 10.12,
10.34\tabularnewline
\hline 
2 substitution/site & 5.66, 5.7b, 5.11a, 5.11b, 5.11c, 5.16, 6.7a, 6.7b, 6.17a, 6.17b\tabularnewline
\hline 
3 substitution/site & 4.4b, 5.7c\tabularnewline
\hline 
>3 substitution/site & 3.4, 4.5a, 4.5b\tabularnewline
\hline 
never & 2.2b, 3.3a, 3.3b, 3.3c, 4.4a\tabularnewline
\hline 
\end{tabular}\hfill{}

\caption{Approximate levels of saturation of model Markov matrices before their
product matrix has significant (>5\%) chance of being non-embeddable
(i.e. ``average'' rate matrix $\widehat{Q}$, as defined in the text,
is non-stochastic). Data derived from Monte Carlo simulation. \label{tab:nonstochastic}}

\end{table}

Thus, we see the ``local'' in the ``local multiplicative closure'' of Lie Markov models is really quite broad: phylogenies have to be quite deep before non-embeddability potentially becomes an issue, and very deep before the average $Q$ becomes complex. 
Under most practical circumstances where we would be attempting to reconstruct phylogenies from real data, the Lie Markov models can safely be considered to be simply ``multiplicatively closed'', without further reference to the ``local'' condition.

It is natural to expect that the more different $Q_{1}$ and $Q_{2}$ are, the more likely it is that $Q$ will be non-stochastic. 
We tested this on models 6.6, 8.8, 8.10b and 10.12. 
Using a trace value for $(Q_{1},Q_{2})$ which resulted non-embeddability rate close to 50\%, we generated a thousand random $(Q_{1},Q_{2})$ pairs, then measured the difference $|Q_{1}-Q_{2}|$ (where $|\ldots|$ indicates the root mean square of off-diagonal elements). The mean difference for non-embeddable pairs was higher than for embeddable pairs, but only by about 0.3 standard deviations, so embeddability is only weakly dependent on
the difference between the input rate matrices.

\section{Discussion}

\global\long\def\swrs{\mathfrak{S}_{2}\wr\mathfrak{S}_{2}}

If we model DNA mutation as non-homogeneous across a phylogeny, using
a model which does not have multiplicative closure leads to a lack
of consistency \citep{sumner2012b}. With such a model, applying
a single set of model parameters to a given edge cannot reproduce
the effects of model parameters varying with time along that edge.
The Lie Markov models were developed to avoid this problem \citep{LieMarkov}.
The fully symmetric Lie Markov models are few in number (1.1 (JC),
3.3a (K3ST), 4.4a (F81) 6.7a (K3ST+F81), 9.20b (doubly stochastic) and 12.12 (GM)).
By relaxing the symmetry condition to allow one pairing of DNA bases
to be distinguished, we greatly increase the number of available models
whilst also allowing for the transition/transversion (RY) distinction
which is common in DNA models (e.g. K2ST, HKY). 
We call the Lie Markov models which allow for the RY distinction the RY Lie Markov models, although we include within this category the models which distinguish the WS and MK base pairings also.

A classification of the RY Lie Markov models  was derived in \citet{fernandez2014}, with emphasize on the mathematical derivation and structure of the models. 
In addition to the fully symmetric Lie Markov models, a further 32 Lie Markov
models were found to exist, most of which are novel.  
In this paper we have presented the models in a more accessible way, explored their applicability to real data sets, and dealt with implementation issues around how to parameterize the models.
For the 31 useful RY Lie Markov models, we also considered allowing alternative base pairs to be distinguished: the WS pairing and the MK pairing. 
The WS pairing is more natural to consider than RY for sequences where there is no distinction between the DNA strands, as is usually the case for non-coding DNA. 

We compared the performance of the Lie Markov models to the standard benchmark of the GTR model and popular submodels.
The majority of Lie Markov models are not time reversible, but we argue that in the context of a non-homogeneous mutation process, time reversibility has already been lost, so, beyond algorithmic details, this is not a modeling disadvantage.

We tested the models on a diverse set of Eukaryotic DNA data sets.
For each data set, we fixed the tree topology and then optimized the log likelihood over model parameters and branch lengths. 
The optimal log likelihoods of the models were compared via the Bayesian information
criterion. 
A selection of more traditional time reversible models were included in the analysis for purposes of comparison. 
The results show that the RY Lie Markov models are biologically plausible, with five of the seven data sets selecting a Lie Markov model as the optimal model (although in one case, the model is the previously studied General Markov model). 
One data set (of buttercup chloroplast mostly intergenic DNA) stood out from the rest as strongly favouring the MK Lie Markov models.

We have shown how the basis matrix structure of the RY Lie Markov models determines the nesting relationships of the models, and the equilibrium base frequencies that the models can generate. 
Additionally, when implementing the Lie Markov models, the problem of parameterizing the space of stochastic rate matrices is non-trivial. 
We have presented a parameteriziation which successfully achieves this, with relative simplicity. 

The theoretical results of \cite{LieMarkov} and \cite{fernandez2014} prove only that the Lie Markov models have ``local multiplicative closure''. 
This means that the ``average'' rate matrix of a time varying process can be non-stochastic or even complex. 
We performed some Monte Carlo simulations to conclude that multiplicative closure (i.e. a real, stochastic average rate matrix) is very likely to be maintained in all phylogenetic analyses excepting those with very deep divergences (for which, as sequences are nearly uncorrelated across deep divergence, the choice of model is not very important anyhow).

Our future plans include testing the models in a non-time-homogeneous context, performing likelihood analysis on many more data sets, and expanding the range of software which implements the models.

\section{Software}

The program used to generate data for Tables~\ref{tab:BIC} and \ref{tab:BIC ranks} was written in Java and uses a modified version of the PAL library \citet{PAL}. 
It is available on request from the lead author, but we caution that this is research rather than production software, not designed for ease of use or robustness.
We are in the process of coding a reference implementation of the Lie Markov models as a Beast2 plugin. 
Should the reader wish to implement the models in your own software, we are happy to assist.

\section*{Supplementary Material}

Supplementary tables contain a more complete listing of BIC values and ranking of models, and independent reparameterizations for each model. 
Additional supplementary files are a spreadsheet of all the likelihood values plus AICc as well as BIC calculations, a Mathematica notebook which derives the independent reparameterizations, the Java code used to calculate the likelihoods, and Beast2 plug-in code.

\bibliographystyle{abbrvnat}
\bibliography{lm}

\end{document}